# Faster hyperspectral image classification based on selective kernel mechanism using deep convolutional networks

Guandong Li, Chunju Zhang

*Abstract*- Hyperspectral imagery is rich in spatial and spectral information. Using 3D-CNN can simultaneously acquire features of spatial and spectral dimensions to facilitate classification of features, but hyperspectral image information spectral dimensional information redundancy. The use of continuous 3D-CNN will result in a high amount of parameters, and the computational power requirements of the device are high, and the training takes too long. This letter designed the Faster selective kernel mechanism network (FSKNet), FSKNet can balance this problem. It designs 3D-CNN and 2D-CNN conversion modules, using 3D-CNN to complete feature extraction while reducing the dimensionality of spatial and spectrum. However, such a model is not lightweight enough. In the converted 2D-CNN, a selective kernel mechanism is proposed, which allows each neuron to adjust the receptive field size based on the two-way input information scale. Under the Selective kernel mechanism, it mainly includes two components, se module and variable convolution. Se acquires channel dimensional attention and variable convolution to obtain spatial dimension deformation information of ground objects. The model is more accurate, faster, and less computationally intensive. FSKNet achieves high accuracy on the IN, UP, Salinas, and Botswana data sets with very small parameters.

## I. INTRODUCTION

HYPERSPECTRAL images (HSIs) not only reflect the spectral information of the feature, but also includes the geometric structure and spatial distribution of the feature. The type and development of the ground object caused the change of the value of the remote sensing image data. By analyzing the variation law of the image data, the type of the ground object can be effectively identified and classified. Accurate classification of HSI is an important task in practical applications in various fields, and has a wide range of applications in agriculture, astronomy, and environmental science.

In the past few decades, the classification of HSI has been very active. Feature extraction means include artificially designed feature extraction techniques and learning-based feature extraction techniques. The use of hand-designed feature descriptions[2] includes several HSI classification methods. [3] - [4] propose sparse representations or patch-based sparse representations as spatial features. A composite kernel is used to combine spatial and spectral information for HSI classification[5]. Later, the academic community proposed a number of spatial-spectral correlation models [6] - [7] to improve classification accuracy. In [8], based on the spatial information of the SVM classifier, multi-scale superpixel segmentation is used to model the distribution of the class. In [9], three kernels are used to take advantage of spatial spectral information and combine them for classification.

In recent years, deep learning technology has made great progress in the field of remote sensing research, especially convolutional neural networks have great advantages over manual extraction features. Zhao et al. [10] proposed using principal component analysis to reduce the original hyperspectral image, using 2D-CNN to extract spatial features on the reduced-dimensional image, and using balanced local discriminant embedding to extract spectral Information, and finally combine spatial and spectral information into the classifier to improve classification accuracy. Chen et al. [11] proposed a hyperspectral feature extraction framework based on CNN, and discussed the extraction effects of 1D-CNN, 2D-CNN and 3D-CNN. Based on 3D-CNN, Zhong et al. [12] proposed using the supervised spatial-spectral residual network SSRN, design spatial and spectral residual module to extract spatial and spectral information respectively, which is a powerful extension of 3D-CNN using residual structure. Li et al. [13] proposed the use of 3D-CNN combined with DenseNet's dense connection structure to achieve deep extraction of HSI features. [14] proposed a dual path network combining residual and dense connections. [15] proposed HybirdSN to explore the feature extraction method of 3D-2D CNN for hyperspectral image classification, but the accuracy is general and the model parameters are very high.

It is obvious from the literature that some of the methods used for classification only use 2D-CNN, do not make full use of the correlation between the spatial-spectral information, and some use 3D-CNN, simultaneously sampling in the spatial and spectral dimensions to obtain good accuracy. However, the HSI information spectral dimension information is redundant. The use of continuous 3D-CNN will generate a lot of parameter quantities, and the computational power requirement for the device is high, and the training takes too long. This letter is used to solve the above two problems. First, combining 3D-CNN and 2D-CNN, using 3D-CNN to complete feature extraction while reducing the dimensionality of spatial and spectral. Secondly, whether it is based on 3D-CNN or 2D-CNN models, it tends to deepen the network structure and contribute to deep feature extraction. But such a model is not lightweight enough. We proposed a selective kernel mechanism in the converted 2D-CNN, which allows each neuron to adjust the receptive field size based on two-way input information scale adaptively. Under the Selective kernel mechanism, it mainly includes two components, SE module [16] and deformable convolution [17]. SE acquires channel dimensional attention and deformable convolution to obtain spatial dimension ground object deformation information. Feature extraction is performed spatially using deformable convolutions with two inconsistent kernel sizes. Two-way information is passed through the SE module to achieve spatial and dimensional feature association. The selective kernel mechanism dynamically selects the region of the receptive field and attention, and maximizes the extraction of useful information with higher accuracy, faster speed, and less computational power. We call the Faster selective kernel mechanism network (FSKNet). FSKNet achieves high accuracy on the IN, UP, Salinas, and Botswana data sets with very small parameter quantities.

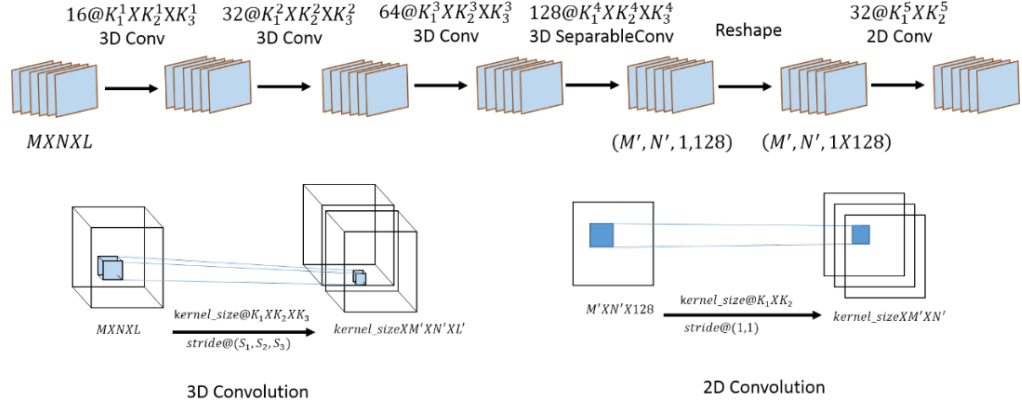

Fig. 1. Combining 3D-CNN and 2D-CNN conversion module

## II. PROPOSED METHOD

### 2.1 Combining 3D-CNN and 2D-CNN conversion modules

For HSI classification, 2D-CNN only applies convolution on the spatial dimension, however we prefer to capture spectral information encoded in multiple frequency bands and spatial information. The 3D-CNN kernel can extract band and spatial feature representations from HSI, but at the cost of increased computational complexity and the model becomes very slow. In order to make full use of the 2D and 3D CNN automatic learning features, we propose a hybrid feature learning module combining 3D CNN and 2D CNN. HSI spectral dimension information redundancy, in order to eliminate spectral redundancy, first apply the traditional PCA to the original HSI data along the spectral band. PCA reduces the number of spectral bands while maintaining the same spatial size, retaining more important information, but PCA also suffers loss of information while performing dimensionality reduction. We believe that the ideal dimension reduction method should be to reduce the size of the data input to the 3d cube while extracting the spatial and spectral features. As shown in Figure x, the first three layers are 3D stride convolution layers. At the same time of spatial dimensionality reduction, we chose a larger dimensionality reduction dimension in the spectral dimension, so that the spatial and spectral simultaneous sampling and synchronous dimensionality reduction are realized in the 3D-CNN framework. Unlike PCA, which reduces the dimension before input, our dimension reduction method is hidden in feature extraction. As the convolution layer increases, the feature map becomes smaller, but the number of filters growssignificantly. In order to reduce the amount of parameters, we use separable convolution in the fourth layer. The most important part of the conversion mechanism is the Reshape layer. Since each feature vector is convolved from the same raw data with different convolution kernels, which makes these vectors have different representations of particular features, there is a strong correlation between these vectors. Reshape all original vectors and integrate channel and spectral dimensions. After the four-layer convolution, the spectral dimension has been reduced to a small value (we reduced the spectral dimension to 1 by controlling the stride size), and we can think that the model has learned a wealth of spectral knowledge, at this time (shape[1], shape [2], shape[3], shape[4]) becomes (shape[1], shape[2], shape[3]xshape[4]). After the reshaping operation, the data can be input to our selective kernel module as a normal 2D image classification, thereby converting the 3D data to the 2D data in this way. This conversion mechanism can achieve good performance and avoid excessive Fitting.

### 2.2 Selective kernel mechanism

After using the combination of 3D-CNN and 2D-CNN conversion modules, we get 2D input image. In order to enable the neurons to be adaptive to the size of the receptive field, we proposed selective kernel mechanism for automatic selection operations in two kernel branches with different kernel sizes. The mechanism consists of two major structures. The first structure consists of two deformable convolutions with different kernel sizes. The use of deformable convolution is more effective in spatially efficient extraction of spectral images. From the conversion of 3D to 2D-CNN, there is no loss of spatial information. The use of two different branches of the kernel is to allow the model to automatically adjust the regional receptive field. This can be used to deal with the characteristics of sparse features in HSI. In order to keep the output feature size of the first structure consistent, we used a detailed convolution kernel in the 5x5 kernel branch. The second structure is a SE module, a lightweight gating mechanism designed to simulate channel relationships in a computationally efficient manner. It is designed to enhance the representation of basic modules throughout the network. Applying an attention mechanism to HSI, it biases the allocation of valid sample resources available for processing to the region with the richest input signal. It accelerates the understanding of the type of features by the deep learning model and facilitates the final classification. The SE module can suppress invalid information, activate valid information, and pass back the weighted output information. The SE module consists of two operations, the first is squeeze, the global information is compressed into a channel descriptor using globalaveragepooooling, the second is excitation, adaptive recalibration, using the sigmoid function to obtain the normalized weight between 0-1, and then normalized weights are weighted to the characteristics of each channel by a scale operation. These two branches get a feature map of the mixed receptive field, and the network can adaptively adjust the receptive field size according to the learned data content. We use this gate mechanism to control the flow of information from two branches carrying different information scales to the next neuron.

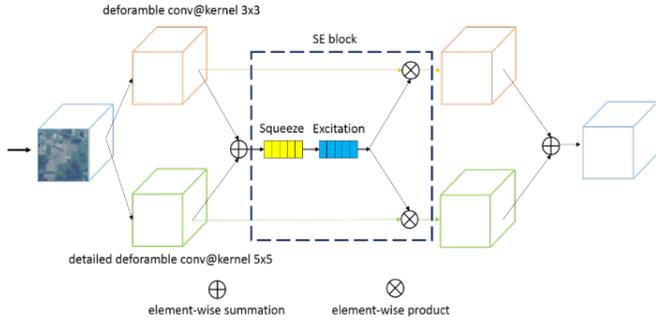

Fig. 2. Selective kernel mechanism

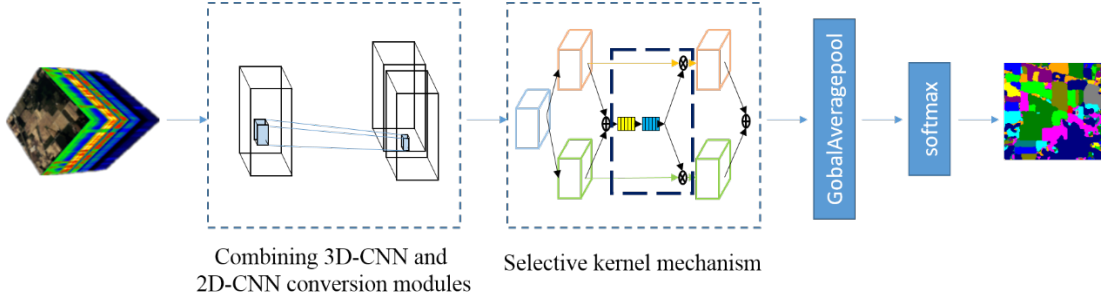

Fig. 3. Illustration of the architecture of the FSKNet

2.3 Network architecture

The overall network structure is as shown in the figure. Input 3D HSI, obtain 2D image through conversion layer, complete feature extraction and dimensionality reduction of spatial spectral dimension, and then perform multi-scale feature correlation through the selective kernel module to dynamically select the kernel. In the structure, you can add multiple selective kernel modules to enhance feature association. But in fact, on the four HSI datasets tested, a selective kernel can achieve high accuracy, and the parameter of the model is much lower than the state of the art. Finally, through the gobalavgeragepool to get the input softmax vector, we do not use the fc layer, fc parameter is high, global pooling can be a good substitute for fc. Table I is a structural table of the model on the IN data set.

TABLE I STRUCTURE TABLE OF THE MODEL ON THE IN DATA SET

| Layers | Output Size | Connected to | Param |
| --- | --- | --- | --- |
| input_1 (InputLayer) | (None, 19, 19, 200, 1) | | 0 |
| conv3d_1 (Conv3D) | (None, 17, 17, 28, 16) | input_1 | 1008 |
| batch_normalization_1 (BatchNormalization) | (None, 17, 17, 28, 16) | conv3d_1 | 64 |
| conv3d_2 (Conv3D) | (None, 15, 15, 5, 32) | batch_normalization_1 | 23040 |
| batch_normalization_2 (BatchNormalization) | (None, 15, 15, 5, 32) | conv3d_2 | 128 |
| conv3d_3 (Conv3D) | (None, 13, 13, 1, 64) | batch_normalization_2 | 55296 |
| batch_normalization_3 (BatchNormalization) | (None, 13, 13, 1, 64) | conv3d_3 | 256 |
| separable_conv3d_1 (SeparableConv3D) | (None, 11, 11, 1, 128) | batch_normalization_3 | 8768 |
| reshape_1 (Reshape) | (None, 11, 11, 128) | separable_conv3d_1 | 0 |
| conv2d_1 (Conv2D) | (None, 11, 11, 32) | reshape_1 | 4096 |
| batch_normalization_4 (BatchNormalization) | (None, 11, 11, 32) | conv2d_1 | 128 |
| deformableconv _1 (Deformableconv2D) | (None, 11, 11, 64) | batch_normalization_4 | 36864 |
| deformableconv _2 (Deformableconv2D) | (None, 11, 11, 64) | batch_normalization_4 | 69632 |
| batch_normalization_3 (BatchNormalization) | (None, 11, 11, 64) | deformableconv _1 | 256 |
| batch_normalization_4 (BatchNormalization) | (None, 11, 11, 64) | deformableconv _2 | 256 |
| add_1 (Add) | (None, 11, 11, 64) | batch_normalization_3 batch_normalization_4 | 0 |
| global_average_pooling2d_1 (GlobalAveragePooling2D) | (None, 64) | add_1 | 0 |

| Layer | Output Shape | Connected to | Param # |
|---|---|---|---|
| reshape_2 (Reshape) | (None, 1, 1, 64) | global_average_pooling2d_1 | 0 |
| dense_1 (Dense) | (None, 1, 1, 4) | reshape_2 | 256 |
| dense_2 (Dense) | (None, 1, 1, 64) | dense_1 | 256 |
| multiply_1 (Multiply) | (None, 11, 11, 64) | batch_normalization_5 dense_2 | 0 |
| multiply_2 (Multiply) | (None, 11, 11, 64) | batch_normalization_6dense_2 | 0 |
| add_2 (Add) | (None, 11, 11, 64) | multiply_1/multiply_2 | 0 |
| separable_conv2d_1 (SeparableConv2D) | (None, 9, 9, 64) | add_2 | 4672 |
| separable_conv2d_2 (SeparableConv2D) | (None, 7, 7, 128) | separable_conv2d_1 | 8768 |
| global_average_pooling2d_2 (GlobalAveragePooling2D) | (None, 128) | separable_conv2d_2 | 0 |
| dense_3 (Dense) | (None, 16) | global_average_pooling2d_2 | 2064 |

Total params: 215,808
Trainable params: 215,264
Non-trainable params: 544

## III EXPERIMENTS

### 3.1 Datasets

The first data set is the Indian Pines dataset. It was collected in June 1992 by the AVIRIS spectral imager at the Indiana sPine Forest Experimental Area in Northwest Indiana. It measures 145 x 145 pixels with a spatial resolution of 20 meters and a wavelength range of 0.4-2.5 microns for the 220 band.

The second data set is Pavia University dataset. It was collected by the ROSIS spectral imager in 2001 in the Pavia region of northern Italy. The image size is 610 × 340 pixels, thespatial resolution is 13 m, contains 115 bands, and the wavelength range is 0.43-0.86 μm for the 103 band.

The third data set is Salinas dataset. It was collected by AVIRIS sensors located in the Salinas Valley, California, with 224 bands and a high spatial resolution of 3.7 m pixels. The pixel space size is 512×217 pixels.

The fourth data set is the Botswana dataset. The NASA EO-1 satellite acquired a series of data from the Okavango Delta in Botswana from 2001 to 2004. The Hyperion sensor on EO-1 acquires data of 30 m pixel resolution over a 7.7 km strip covering 242 bands of the 400-2500 nm spectral portion in a 10 nm window.

In our experiments, we used OA, AA and Kappa as the main indicators to evaluate the accuracy of the model. We used training and test time, flops as the main indicator to evaluate the speed of the model.

### 3.2 Discussion parameters

3.2.1The effect of the size of neighboring pixel blocks on the classification accuracy of the model neighboring pixel blocks are important factors in the size of the HSI input into the model. We discussed the changes in OA, AA, and Kappa for FSKNet on four datasets when changing from 15-23, as shown in the table. On the IN, UP, Botswana and Salinas dataset, the neighboring pixel blocks were selected to be 23, 15, 15, and 17.Due to hardware limitations, only the 15 and 17 groups were selected on the salinas dataset. The training set ratio of FSKNet is 5:1:4.

TABLE II INFLUENCE OF NEIGHOURHOOD PIXEL BLOCKS ON THE PRECISION OF IN, UP, SALINAS AND BOTWANA DATA SETS

| Neignboring pixel | IN | | | UP | | | Botswana | | | Salinas | | |
|---|---|---|---|---|---|---|---|---|---|---|---|---|
| | OA | AA | Kappa | OA | AA | Kappa | OA | AA | Kappa | OA | AA | Kappa |
| 15 | 99.58 | 99.20 | 99.53 | 99.96 | 99.94 | 99.95 | 1 | 1 | 1 | 99.85 | 99.70 | 99.84 |
| 17 | 99.68 | 99.27 | 99.64 | 99.96 | 99.89 | 99.94 | 1 | 1 | 1 | 99.94 | 99.94 | 99.94 |
| 19 | 99.63 | 99.22 | 99.58 | 99.88 | 99.70 | 99.85 | 1 | 1 | 1 | - | - | - |
| 21 | 99.78 | 99.68 | 99.75 | 99.92 | 99.84 | 99.89 | 1 | 1 | 1 | - | - | - |
| 23 | 99.83 | 99.73 | 99.81 | 99.81 | 99.57 | 99.74 | 99.85 | 99.86 | 99.83 | - | - | - |

3.2.2 The effect of the size of training ratio on the classification accuracy of the model

After choosing the best neighboring pixel block parameters, we discussed the impact of the training set size. On the IN and UP datasets, the accuracy is preferably 5:1:4. In the Salinas dataset, the highest accuracy is achieved when the training set ratio is 4:1:5. In Botswana dataset, when the training set ratio is 3:1:6, the accuracy is 1., and the dataset is also the most prominent data set of FSKNet.

TABLE III THE EFFECT OF TRAINING SET SCALE ON PRECISION ON IN, UP, SALINAS, BOTWANA DATA SETS

| Training ratio | IN | | | UP | | | Botswana | | | Salinas | | |
|---|---|---|---|---|---|---|---|---|---|---|---|---|
| | OA | AA | Kappa | OA | AA | Kappa | OA | AA | Kappa | OA | AA | Kappa |
| 2:1:7 | 99.34 | 99.22 | 99.25 | 99.90 | 99.89 | 99.87 | 99.34 | 99.43 | 99.28 | 99.90 | 99.87 | 99.89 |
| 3:1:6 | 99.48 | 99.42 | 99.41 | 99.93 | 99.91 | 99.91 | 1 | 1 | 1 | 99.86 | 99.73 | 99.85 |
| 4:1:5 | 99.63 | 99.08 | 99.58 | 99.80 | 99.68 | 99.73 | 1 | 1 | 1 | 99.98 | 99.98 | 99.98 |
| 5:1:4 | 99.83 | 99.73 | 99.81 | 99.96 | 99.94 | 99.95 | 1 | 1 | 1 | 99.94 | 99.94 | 99.94 |
| 6:1:3 | 99.71 | 99.41 | 99.67 | 99.90 | 99.91 | 99.87 | 1 | 1 | 1 | - | - | - |

### 3.3 Experimenal Results

In Table IV, the FSKNet parameter quantities and the required flops are significantly lower than the comparison model.In Tables V and VI, we compared the five indicators of OA, AA, Kappa, training time and Test time. FSKNet has achieved a big advantage in all four data sets. On the IN dataset, all five indicators are optimal. On the UP dataset, only the test time is slightly slower. In the Salinas dataset, although the accuracy is behind the SSRN and 3D-DenseNet, the training time and test time are the shortest, especially compared to 3D-DenseNet. The accuracy is optimal on the Botwana dataset.

TABLE IV COMPARISON OF PARAMETER QUANTITIES AND FLOPS OF 5 MODELS

| | 3D-CNN | SSRN | 3D-DenseNet | HybridSN | FSKNet |
|---|---|---|---|---|---|
| Params | 16394652 | 346788 | 2562452 | 5503108 | 215812 |
| Flops | 81959692 | 1732909 | 5234500 | 11005179 | 893584 |

TABLE V COMPARISON OF 5 METHODS ON IN AND UP DATA SETS

| Methods | IN | | | | | UP | | | | |
|---|---|---|---|---|---|---|---|---|---|---|
|  | OA | AA | Kappa | Training time | Test time | OA | AA | Kappa | Training time | Test time |
| 3D-CNN | 99.61 | 98.63 | 99.55 | 13884.72 | 16.09 | 99.41 | 99.37 | 99.22 | 4203.72 | 6.25 |
| SSRN | 99.56 | 93.06 | 99.50 | 6336.10 | 11.10 | 99.88 | 99.76 | 99.85 | 8389.92 | 10.98 |
| 3D-DenseNet | 99.95 | 99.66 | 99.94 | 32232.29 | 43.66 | 1 | 1 | 1 | 37216.36 | 45.81 |
| HybridSN | 99.83 | 99.30 | 99.81 | 5707.83 | 8.73 | 99.80 | 99.69 | 99.73 | 5312.18 | 9.58 |
| FSKNet | 99.83 | 99.73 | 99.81 | 2995.59 | 6.99 | 99.96 | 99.94 | 99.95 | 7242.29 | 10.75 |

TABLE VI COMPARISON OF 5 METHODS ON SALINAS AND BOTWANA DATASETS

| Methods | Salinas | | | | | Botswana | | | | |
|---|---|---|---|---|---|---|---|---|---|---|
|  | OA | AA | Kappa | Training time | Test time | OA | AA | Kappa | Training time | Test time |
| 3D-CNN | 99.36 | 98.86 | 99.29 | 30236.66 | 1200.12 | 99.90 | 99.91 | 99.89 | 725.83 | 1.81 |
| SSRN | 99.99 | 99.99 | 99.99 | 17661.24 | 116.19 | 1 | 1 | 1 | 522.52 | 1.64 |
| 3D-DenseNet | 99.99 | 99.99 | 99.99 | 78055.12 | 193.26 | 99.64 | 99.60 | 99.61 | 2289.37 | 6.62 |
| HybridSN | 99.87 | 99.94 | 99.86 | 14217.54 | 110.41 | 99.95 | 99.95 | 99.94 | 351.98 | 1.25 |
| FSKNet | 99.98 | 99.98 | 99.98 | 13420.11 | 108.39 | 1 | 1 | 1 | 424.82 | 1.29 |

## IV CONCLUSION

In this letter, we present the current real-time network structure for processing HSIs. Based on the 3D-2D CNN conversion module and the selective kernel mechanism, FSKNet completes the spectral dimension reduction when the spatial spectral features are jointly extracted. The selective kernel module allows each neuron to adapt to the field size based on two-way input information. FSKNet maximizes the extraction of useful information with higher precision, faster speed and less computational power. It achieves state of the art with excellent precision and speed.